\documentclass[12pt]{article}

\usepackage{fullpage}
\usepackage{amsmath, amssymb, hyperref}

\begin{document}

\title{Novel counterexample to the Nelson-Seiberg theorem}
\author{James Brister\textsuperscript{*},
        Zheng Sun\textsuperscript{\dag}\\
        \normalsize\textit{College of Physics, Sichuan University,}\\
        \normalsize\textit{29 Wangjiang Road, Chengdu, Sichuan, 610064, PRC}\\
        \normalsize\textit{E-mail:}
        \textsuperscript{*}\texttt{jbrister@scu.edu.cn,}
        \textsuperscript{\dag}\texttt{sun\_ctp@scu.edu.cn}
       }
\date{}
\maketitle

\begin{abstract}
We present a new type of counterexample to the Nelson-Seiberg theorem.  It is a generic R-symmetric Wess-Zumino model with nine chiral superfields, including one field of R-charge $2$ and no R-charge $0$ field.  As in previous counterexamples, the model gives a set of degenerate supersymmetric vacua with a non-zero expectation value for a pair of oppositely R-charged fields.  However, one of these fields appears quadratically in the superpotential, and many other fields with non-zero R-charges gain non-zero expectation values at the vacuum, and so this model escapes the sufficient condition for counterexamples established in previous literature.  Thus there are still open problems in the relation of R-symmetries to supersymmetry breaking in generic models.
\end{abstract}

\section{Introduction}

The Nelson-Seiberg theorem relates R-symmetries to superymmetry (SUSY) breaking in generic $\mathcal{N} = 1$ Wess-Zumino models.  The original result~\cite{Nelson:1993nf} states that the presence of an R-symmetry is a necessary condition, and a broken R-symmetry is a sufficient condition, for SUSY breaking at the stable vacuum of a generic model.  A refinement of this result~\cite{Kang:2012fn, Li:2020wdk} relates the existence of a SUSY vacuum to the numbers of fields with certain R-charges in a model with a polynomial superpotential.  However, exceptions~\cite{Sun:2019bnd} to both of these results have been found, in which a model with generic coefficients breaks the R-symmetry at the SUSY vacuum.  The source of these exceptions has been identified~\cite{Amariti:2020lvx} as pairs of fields with opposite R-charges obtaining vacuum expectation values (VEVs).  Features of these exceptions can be summarized into a sufficient condition~\cite{Sun:2021svm}.  To summarize: a sufficient condition for the existence of a SUSY vacuum in a generic R-symmetric Wess-Zumino model is that the number of R-charge $2$ fields is less than or equal to the sum of the number of R-charge $0$ fields and the number of independent products of oppositely R-charged fields which appear only linearly in cubic terms of a renormalizable superpotential.

In this note, we demonstrate that this sufficient condition is not also necessary, by constructing a generic R-symmetric superpotential which does not satisfy the above condition.  The model nonetheless possesses a set of SUSY vacua where many fields with non-zero R-charges gain non-zero VEVs.  Therefore this model is a counterexample to the Nelson-Seiberg theorem, and escapes the sufficient condition established in previous literature.

The rest of this paper is arranged as follows. Section 2 reviews the sufficient condition for SUSY vacua in R-symmetric Wess-Zumino models which covers all previous counterexamples.  Section 3 presents the new counterexample and its vacuum structure, showing that it is a counterexample escaping the previous sufficient condition. Section 4 discusses properties of the SUSY vacuum and implications of the result.

\section{The sufficient condition for SUSY vacua}

Here we briefly summarize the results of \cite{Kang:2012fn, Sun:2021svm}; for details, we refer readers to those papers.

Under a continuous $U(1)$ R-symmetry, where the R-charge for Grassmann numbers $\theta^\alpha$ is set to $1$, the superpotential $W(\phi_i)$, built from scalar fields $\phi_i$ or their corresponding chiral superfields, must have R-charge $2$ to make the SUSY action R-invariant.  Thus only R-charge $2$ fields may appear as linear terms in the superpotential.  Following the convention of \cite{Sun:2021svm}, we call such fields $X_i$.  The terms linear in $X_i$ which may appear in a renormalizable superpotential are
\begin{equation}
W_X = a_i X_i
      + b_{i j} X_i Y_j
      + c_{i j k} X_i Y_j Y_k
      + d_{(r) i j k} X_i P_{(r) j} Q_{(- r) k},
\end{equation}
where $a_i$, $b_{i j}$, $c_{i j k}$ and $d_{(r) i j k}$ are coefficients, $Y_j$ are R-charge $0$ fields, and the fields $P_{(r) i}$ and $Q_{(- r) i}$ have opposite R-charges $\pm r$, so that their product is R-neutral.  In addition, the assumption is made that the $P$ and $Q$ fields appear only linearly in cubic terms.  Thus in addition to $W_X$, other terms which may appear in a renormalizable superpotential are
\begin{equation}
\begin{split}
W_A &= \underbrace{\xi_{i j k} X_i X_j A_k}_{r_k = - 2}
       + \underbrace{\rho_{i j k} X_i A_j A_k}_{r_j + r_k = 0}
       + \underbrace{\sigma_{(r) i j k} P_{(r) i} A_j A_k}_{r_j + r_k = 2 - r}
       + \underbrace{\tau_{(r) i j k} Q_{(- r) i} A_j A_k}_{r_j + r_k = 2 + r}\\
    &\quad
       + \underbrace{(\mu_{i j} + \nu_{i j k} Y_k) A_i A_j}_{r_i + r_j = 2}
       + \underbrace{\lambda_{i j k} A_i A_j A_k}_{r_i + r_j + r_k = 2},
\end{split}
\end{equation}
where $\xi_{i j k}$, $\rho_{i j k}$, $\sigma_{(r) i j k}$, $\tau_{(r) i j k}$, $\mu_{i j}$, $\nu_{i j k}$ and $\lambda_{i j k}$ are coefficients, and $A_i$ are fields which have R-charges not equal to $2$ or $0$ and can not be identified as $P$ or $Q$ fields.  The full superpotential
\begin{equation}
W = W_X + W_A
\end{equation}
contains all possible R-charge $2$ terms built from all fields in our classification according to their R-charges.

When seeking SUSY vacua, that is, solutions to the F-term equations 
\begin{equation}
\partial_i W = \frac{\partial W}{\partial \phi_i}
             = 0,
\end{equation}
one can satisfy all the F-term equations coming from derivatives with respect to $Y$, $P$, $Q$ and $A$ fields, by assuming that only $Y$, $P$ and $Q$ fields obtain non-zero VEVs.  The number of F-term equations coming from derivatives with respect to $X$ fields is equal to $N_X$, the number of $X$ fields, while the number of independent variables in these equations is equal to the sum of $N_Y$, the number of $Y$ fields, plus $N_{P Q}$, the number of independent $P$-$Q$ pair products, which can be expressed as
\begin{equation}
N_{P Q} = \sum_r \left ( N_{P(r)} + N_{Q(- r)} - 1 \right ),
\end{equation}
where $N_{P(r)}$ and $N_{Q(- r)}$ are the numbers of $P$ and $Q$ fields with R-charges $\pm r$ and the sum is taken only over values of $r$ for which $N_{P(r)}$ and $N_{Q(- r)}$ are non-zero.  These equations are always solvable~\cite{Li:2021ydn} for generic superpotential coefficients if the number of equations is less than or equal to the number of variables, and so a sufficient condition for the existence of SUSY vacua is
\begin{equation} \label{eq:2-01}
N_X \le N_Y + N_{P Q}.
\end{equation}
This condition includes the case $N_X \le N_Y$, under which the revised Nelson-Seiberg theorem predicts the existence of SUSY vacua~\cite{Sun:2011fq}, and the case $N_Y < N_X \le N_Y + N_{P Q}$ which is satisfied by all previous counterexample models~\cite{Sun:2019bnd, Amariti:2020lvx, Sun:2021svm}.  In the latter case, the facts $N_X > N_Y$ and that $P$ and $Q$ fields get non-zero VEVs for generic superpotential coefficients indicate that models in this case are counterexamples to both the original Nelson-Seiberg theorem~\cite{Nelson:1993nf} and its revison~\cite{Kang:2012fn}.

In the following section, we shall demonstrate a counterexample which does not satisfy the sufficient condition~\eqref{eq:2-01}.  The model gives a set of SUSY vacua where many fields other than $Y$, $P$ and $Q$ fields get VEVs.  The existence of such a new counterexample means that the sufficient condition presented here is not also a necessary condition for SUSY vacua in R-symmetric Wess-Zumino models.

\section{The new counterexample}

Consider a Wess-Zumino model with nine fields: $X$, $B$, $C$, $\Xi_1$, $\Xi_2$, $\Xi_3$, $A_1$, $A_2$ and $A_3$.  The superpotential is given as
\begin{equation}
\begin{split}
W &= X (a + b B C)
    + \Xi_1 (\alpha_1 A_1 + \beta_1 B^2)
    + \Xi_2 (\alpha_2 A_2 + \beta_2 A_3^2)\\
  &\quad
    + \Xi_3 (\alpha_3 B +\beta_3 A_2^2 + \gamma_3 A_1 C)
    + \gamma_1 \Xi_1^2 A_3,
\end{split}
\end{equation}
where $a$, $b$, $\alpha_i$, $\beta_i$, $\gamma_i$ are coefficients.  This superpotential possesses a $U(1)$ R-symmetry, under which the fields have the R-charge assignment:
\begin{equation} \label{eq:3-01}
\{ r_X, r_B, r_C, r_{\Xi_1}, r_{\Xi_2}, r_{\Xi_3}, r_{A_1}, r_{A_2}, r_{A_3} \}
= \{ 2, \frac{8}{15}, - \frac{8}{15}, \frac{14}{15}, \frac{26}{15}, \frac{22}{15}, \frac{16}{15}, \frac{4}{15}, \frac{2}{15} \}.
\end{equation}
This assignment is unique, or equivalently~\cite{Komargodski:2009jf}, there is no other continuous symmetry of the model.  The superpotential above contains all renormalizable terms permitted by this R-symmetry, so it is the form of a generic superpotential given the fields and their R-charges.

For generic values of the coefficients, we have a set of SUSY vacua at
\begin{equation} \label{eq:3-02}
\begin{gathered}
X = \Xi_1
  = \Xi_2
  = \Xi_3
  = 0, \quad
B C = - \frac{a}{b}, \quad
A_1 = - \frac{\beta_1}{\alpha_1} B^2,\\
A_2 = \left ( - \frac{1}{\beta_3}
                \left ( \alpha_3 + \frac{a \beta_1 \gamma_3}{b \alpha_1} \right )
                B
      \right )^{\frac{1}{2}}, \quad
A_3 = \left ( - \frac{\alpha_2^2}{\beta_2^2 \beta_3}
                \left ( \alpha_3 + \frac{a \beta_1 \gamma_3}{b \alpha_1} \right )
                B
      \right )^{\frac{1}{4}},
\end{gathered}
\end{equation}
with a one complex dimensional degeneracy parameterized by the non-zero VEV of $B$.  Like any SUSY vacuum in generic R-symmetric models, the vacua have the property that the superpotential vanishes term-by-term~\cite{Brister:2021xca} and satisfies the bound found in~\cite{Dine:2009sw}.  The R-symmetry is spontaneously broken everywhere on the degeneracy by all the non-zero VEVs of $B$, $C$ and $A_i$.  Thus this model is a counterexample to the Nelson-Seiberg theorem.  The model has $N_X = 1$, $N_Y = 0$.  Although $B$ and $C$ have opposite R-charges, they can not be identified as $P$ and $Q$ fields because $B$ appears quadratically in $\beta_1 \Xi_1 B^2$ and in the quadratic term $\alpha_3 \Xi_3 B$.  Therefore we have $N_{P Q} = 0$ and the model escapes the previous sufficient condition \eqref{eq:2-01}.

We may see the full vacuum structure of the model from the scalar potential
\begin{equation}
V = (\partial^i W)^* \partial_i W,
\end{equation}
where a minimal K\"ahler potential is assumed.  Like any R-symmetric polynomial superpotential which does not contain at least one field of R-charge 2 and at least one field of R-charge 0, the scalar potential has a stationary point at the origin of the field space%
\footnote{At a stationary point, we have $\partial_j V = (\partial^i W)^* \partial_i \partial_j W = 0$.  For a polynomial superpotential, these equations may contain a constant term only if the superpotential contains terms of the form $a X + b X Y$, where $X$ necessarily has an R-charge of 2, and $Y$ an R-charge of 0.  This property is independent of any considerations of genericity, assuming $a, b \ne 0$.}%
.  In this case, this point is a saddle.  Numerical searches also indicate that there are several metastable local minima with $\lvert V \rvert > 0$, thus SUSY-breaking.

Finally, we note that other than $\partial_X W$, which is uncharged, all the F-terms $\partial_i W$ have positive R-charges.  This means that~\cite{Ferretti:2007ec, Ferretti:2007rq, Azeyanagi:2012pc, Sun:2018hnk}, under a complexified R-symmetry 
\begin{equation}
\phi \to e^{- r_\phi t} \phi, \quad
t \in \mathbb{R},
\end{equation}
all the non-$X$ F-terms will tend to zero as $t \to + \infty$.  We thus might have a runaway direction as $C \to \infty$ and $B \to 0$.  However, as the complexified R-symmetry also takes all other fields to zero in this limit, it coincides with the large-$C$ limit of the SUSY solution~\eqref{eq:3-02}.

\section{Discussions}

As we have shown, the model presented in this work has a field count satisfying $N_X > N_Y + N_{P Q}$, which is outside of the previous classes of both the R-symmetric SUSY vacua~\cite{Sun:2011fq} and the R-symmetry breaking SUSY vacua covered by the sufficient condition~\cite{Sun:2021svm}.  That the SUSY vacua are R-symmetry breaking also indicates that the model is a counterexample to the original Nelson-Seiberg theorem.  The existence of such a new counterexample suggests that there are still some unexplored corners in the classification of R-symmetric Wess-Zumino models.

Just like any SUSY vacuum in R-symmetric models, the SUSY vacua in the new counterexample give $W = 0$ at the SUSY vacuum~\cite{Brister:2021xca, Dine:2009sw, Kappl:2008ie}, and the supergravity version of the model also gives SUSY vacua with zero vacuum energy.  One may hope to use the supergravity model as a low energy effective description for flux compactification of type IIB string theory~\cite{Grana:2005jc, Douglas:2006es, Blumenhagen:2006ci, Ibanez:2012zz}, and such string constructions of $W = 0$ SUSY vacua~\cite{DeWolfe:2004ns, DeWolfe:2005gy, Dine:2005gz, Palti:2007pm, Kanno:2017nub, Palti:2020qlc, Kanno:2020kxr} serve as the first step toward vacua with small superpotentials~\cite{Demirtas:2019sip}.  But the R-symmetry breaking feature of the vacua means that some complex structure moduli obtain nonzero VEVs, which send the Calabi-Yau manifold away from the R-symmetric point in its moduli space.  It is then unnatural to turn on only R-symmetric fluxes and obtain an R-symmetric effective superpotential from the start.  Thus silimarly to previous counterexample models, the new counterexample here does not contribute to the string landscape of $W = 0$ SUSY vacua if we only consider R-symmetric SUSY vacua~\cite{Dine:2005gz}, or string vacua with enhanced symmetries~\cite{DeWolfe:2004ns, DeWolfe:2005gy}.  It is still an open question whether these counterexamples could be low energy effective models for other string constructions.

\section*{Acknowledgement}

The authors thank Zhenhuan Li for helpful discussions on the numerical search techniques.  This work is supported by the National Natural Science Foundation of China under the grant numbers 12205208 and 11305110.

\end{document}